%Paper: hep-th/9504113
%From: Ofer Aharony <oferah@post.tau.ac.il>
%Date: Sun, 23 Apr 1995 18:52:34 +0300 (IDT)
%Date (revised): Tue, 4 Jul 1995 18:27:52 +0300 (IDT)

%
% More duality symmetries and the flows between them
%
% April 1995
%

\input harvmac

%\draftmode

% Useful abbreviations

\def\Tr{{\rm Tr ~}}
\def\ti{\tilde i}
\def\tq{\tilde q}
\def\tQ{\tilde Q}

\def\tZ{\tilde Z}
\def\tz{\tilde z}
\def\tt{{2 \over 3}}
\def\ot{{1 \over 3}}
\def\tW{\tilde W}
\def\hW{\hat W}
\def\htW{\hat{\tilde W}}
\def\tnf{\tilde{N_f}}
\def\tj{\tilde j}
\def\tk{\tilde k}
\def\tm{\tilde m}
\def\tn{\tilde n}
\def\tB{\tilde B}
\def\ta{\tilde a}
\def\tb{\tilde b}
\def\tc{\tilde c}
\def\td{\tilde d}

% List of Journals
\def\np#1#2#3{Nucl. Phys. {\bf B#1} (#2) #3}
\def\pl#1#2#3{Phys. Lett. {\bf #1B} (#2) #3}

\def\physrev#1#2#3{Phys. Rev. {\bf D#1} (#2) #3}

\def\ijmp#1#2#3{Int. J. Mod. Phys. {\bf A#1} (#2) #3}

% Reference list
\nref\om{C. Montonen and D. Olive, \pl {72}{1977}{117}; P. Goddard,
J. Nuyts, and D. Olive, Nucl. Phys. {\bf B125} (1977) 1}%
\nref\dualnf{C. Vafa and E. Witten, hep-th/9408074, \np{431}{1994}{3}}%
\nref\swi{N. Seiberg and E. Witten, hep-th/9407087, \np{426}{1994}{19}}%
\nref\swii{N. Seiberg and E. Witten, hep-th/9408099, \np{431}{1994}{484}}%
\nref\klyt{A. Klemm, W. Lerche, S. Yankielowicz and S. Theisen,
hep-th/9411048, \pl{344}{1995}{169};
P. C. Argyres and A. E. Faraggi,
``The vacuum structure and spectrum of N=2 supersymmetric $SU(n)$ gauge
theory",
hep-th/9411057, IASSNS-HEP-94-94}%
\nref\is{K. Intriligator and N. Seiberg, hep-th/9408155, \np{431}{1994}{551}}%
\nref\dualstr{A. Sen, hep-th/9402002, \ijmp {9}{1994}{3707}; hep-th/9402032,
\pl {329}{1994}{217}}%
\nref\witstr{E. Witten, ``String Theory Dynamics in Various Dimensions",
hep-th/9503124, IASSNS-HEP-95-18}%
\nref\duals{N. Seiberg,
hep-th/9411149, \np{435}{1995}{129}}%
\nref\sei{N. Seiberg, hep-th/9402044, \physrev{49}{1994}{6857}}%
\nref\dualo{O. Aharony, ``Remarks on Non--Abelian Duality in N=1
Supersymmetric Gauge Theories", hep-th/9502013, TAUP-2232-95, to appear
in Phys. Lett. B}%
\nref\dualis{K. Intriligator and N. Seiberg, ``Duality, Monopoles, Dyons,
Confinement and Oblique Confinement in Supersymmetric $SO(N_c)$ Gauge
Theories", hep-th/9503179, RU-95-3, IASSNS-HEP-95/5}%
\nref\ls{R. G. Leigh and M. J. Strassler, ``Exactly Marginal Operators
and Duality in Four Dimensional N=1 Supersymmetric Gauge Theory",
hep-th/9503121, RU-95-2}%
\nref\dualk{D. Kutasov, ``A Comment on Duality in N=1 Supersymmetric
Non--Abelian Gauge Theories", hep-th/9503086, EFI-95-11}%
\nref\efgr{S. Elitzur, A. Forge, A. Giveon, E. Rabinovici, ``More Results
in N=1 Supersymmetric Gauge Theories", hep-th/9504080, RI-4-95}

% Title page

\Title{hep-th/9504113, TAUP-2246-95, CERN-TH/95-91}
{\vbox{\centerline{Flows and Duality Symmetries in}
\centerline{N=1 Supersymmetric Gauge Theories}}}
\centerline{
O. Aharony\foot{Work supported in part by
the US-Israel Binational Science Foundation and by GIF -- the German-Israeli
Foundation for Scientific
research}$^{,}$\foot{Work
supported in part by the Clore Scholars Programme}
, J. Sonnenschein$^{1,}$\foot{Work
supported in part by the Israel Academy of Science}
}
\vglue .5cm
\centerline{School of Physics and Astronomy}
\centerline{Beverly and Raymond Sackler Faculty of Exact Sciences}
\centerline{Tel--Aviv University}
\centerline{Ramat--Aviv, Tel--Aviv 69978, Israel}
\vglue .5cm
\centerline{and S. Yankielowicz$^{1,}$\foot{On leave of absence from the
School of Physics, Raymond and Beverly Sackler Faculty of Exact Sciences,
Tel--Aviv University}}
\vglue .5cm
\centerline{CERN, Geneva, Switzerland}
\vglue 1cm

\noindent
We present more examples of
dual N=1 SUSY gauge theories. This set of theories is connected by
flows to both Seiberg's and Kutasov's dual theories.
This provides a unifying picture of the various dual theories.
We investigate the dual theories, their flat directions and mass
perturbations.
\vglue.5cm
\noindent CERN-TH/95-91
\Date{April 1995}

% Paper

\newsec{Introduction}

Duality symmetries are playing an increasing role in the understanding of
the behavior of supersymmetric gauge theories at strong coupling. The
electric--magnetic dualities relating strong and weak coupling make
sense in the context of $N=4$ supersymmetric gauge theories
\refs{\om,\dualnf}, where they are expected to be exact symmetries.
They were generalized \refs{\swi,\swii,\klyt} to $N=2$ theories
and to some $N=1$ theories \is\ in
the Coulomb phase, and enabled the computation of non--perturbative
results in these theories. There is growing evidence \refs{\dualstr,
\witstr} that these $SL(2,Z)$ duality symmetries appear in string theory
as well.

A new type of duality symmetry, connecting $N=1$ supersymmetric gauge
theories with (generally) different gauge groups, was suggested by
Seiberg in \duals. This duality relates a standard supersymmetric
QCD (SQCD) theory with gauge group $SU(N_c)$ and $N_f$ flavors of quarks, with
another supersymmetric QCD theory, with gauge group $SU(N_f-N_c)$,
$N_f$ quark flavors and additional gauge--singlet fields with a non--trivial
superpotential coupling the singlets to the dual quarks. The evidence
for Seiberg's duality consists \duals\ of (i) the identification of the global
anomalies and gauge--invariant operators in the chiral rings of the
two theories, (ii) the possibility of flowing (by mass perturbations) between
different pairs of
dual theories and from them to known \sei\ effective descriptions of SQCD (for
small values of $N_f$) in terms of bound states, and (iii) the identification
of all flat directions of the two theories. These include in particular
\dualo\ a flat direction along which both theories flow to the same field
theory in the IR.
The generalization
of this duality to the $SO(N_c)$ case (discussed in \refs{\duals,\dualis}) is
related to the electric--magnetic duality, with the dual quarks interpreted
as monopoles (or dyons) of the original theory. In the $SU(N_c)$ case no
monopoles appear (semi--classically) in the spectrum, so that the connection
of the new duality to the electric--magnetic duality is still obscure,
though there may be some relations between the two \ls. For $N_f=N_c+1$ the
duality goes over to the description of the theory in terms of bound states,
with the dual quarks becoming baryons.

Another duality of this type was suggested by Kutasov \dualk.
The original (``electric") theory in this duality is an $SU(N_c)$ gauge
theory ($N_c > 2$),
with $N_f$ flavors of quarks $Q^i_a$ and anti--quarks $\tQ_{\ti}^a$,
and an additional adjoint field $X$, with a superpotential $W_{el}=\Tr(X^3)$.
The dual (``magnetic") theory is an $SU(2N_f-N_c)$ gauge theory, with $N_f$
flavors of quarks $q_i^a$ and anti--quarks $\tq^{\ti}_a$, an adjoint field
$Y$, and singlet fields $M^i_{\ti}$ (identified with $Q^i \tQ_{\ti}$) and
$N^i_{\ti}$ (identified with $Q^i X \tQ_{\ti}$). The superpotential in
the ``magnetic" theory \dualk\ is given by
$W_{mag} = M^i_{\ti} q_i Y \tq^{\ti} + N^i_{\ti} q_i \tq^{\ti} + \Tr(Y^3)$.
% \eqn\wmag{W_{mag} = M^i_{\ti} q_i Y \tq^{\ti} + N^i_{\ti} q_i \tq^{\ti} +
% \Tr(Y^3).}
The evidence given in \dualk\ for this duality consists of the
identification of the global anomalies and gauge--invariant operators in
the two theories, and the possibility of flowing between different
values of $N_f$ by mass perturbations. As in Seiberg's duality, for
$N_f={1\over 2}(N_c+1)$ the duality becomes a description of the theory
in terms of bound states, with the dual quarks becoming baryons. In other
cases the interpretation of the dual quarks in terms of the
original variables is not clear.

In this paper we present another duality transformation of this type,
which can be reached from Kutasov's duality \dualk\ by flowing along a flat
direction. In addition to the
fields described above, both theories in this duality have $\tnf$ additional
flavors of quarks $Z^j_a$ and anti--quarks $\tZ^a_j$,
which couple to the adjoint field by a coupling of
the form $Z^j_a X^a_b \tZ^b_j$. The
duality takes the gauge group $SU(N_c)$
to $SU(2N_f+\tnf-N_c)$. We shall check the 't~Hooft anomaly conditions and the
identification of gauge--invariant bound states in the two theories, and
analyze several mass perturbations of these theories. We shall also
analyze several flat directions of these theories, finding all of them to be
consistent with the duality.
% This provides support also to the duality
% conjecture of \dualk.
This analysis is non--trivial in view of the appearance of
non--renormalizable terms in the superpotential.
In view of the connection to Kutasov's duality, it provides more support
to the duality conjecture of \dualk.
Our analysis will ignore the possible
quantum corrections to the superpotential in both theories, whose analysis
is left for future work. In most cases we will find a consistent picture
without the quantum corrections. This indicates that in these cases
the quantum corrections do not seem to be important.
Among the theories we flow to we will find Seiberg's dual theories \duals.
Thus, we relate the dualities of Seiberg and Kutasov.
Moreover, some cases which will be considered seem to
indicate a possible relation between the $N=1$ duality and the duality of
$N=2$ theories \refs{\swi-\klyt}.

In section 2 we describe the duality transformation, check the 't Hooft
anomaly matching conditions and identify the gauge--invariant bound states.
In section 3 we analyze flows generated by adding mass perturbations,
and verify that the resulting theories
are consistent with the duality. In section 4 we analyze several flat
directions
of the dual theories, finding them too to be consistent with the duality.
In section 5 we analyze another flat
direction which is more complicated.
This flat direction relates our theories to
Kutasov's dual theories \dualk. We end in section 6 with a summary and
conclusions.

\newsec{The duality transformation}

In this section we present a generalization of the known $N=1$ SUSY
duality \dualk. The
dual theories include, in addition to a field in the adjoint
of the gauge group, two types of quarks and anti--quarks
(each in the fundamental and
anti--fundamental representations of the gauge group). One type of
quark couples to the adjoint field and the other does not. We will have
$N_f$ quarks without, and $\tnf$ quarks
with coupling to the adjoint field.
The duality will take an $SU(N_c)$
gauge theory to an $SU(2N_f+\tnf-N_c)$ gauge theory. The original
(``electric") theory will be asymptotically free whenever
$N_f+\tnf < 2N_c$, and the dual (``magnetic") theory will be asymptotically
free for $3N_f+\tnf > 2N_c$.

\subsec{The ``electric" theory}

The ``electric" theory of the duality we analyze includes two types of quarks.
There are $N_f$ flavors of quarks $Q^i_a$ and anti--quarks $\tQ^a_{\ti}$, with
$a=1,\cdots,N_c$ and $i,\ti=1,\cdots,N_f$, and $\tnf$ flavors of quarks $Z^j_a$
and anti--quarks $\tZ_j^a$, with $a=1,\cdots,N_c$ and $j=1,\cdots,\tnf$.
There is also a field in the adjoint representation of the gauge group
which will be denoted by $X$. We take the superpotential of the
``electric" theory to be, at the classical level,
\eqn\wel{W_{el} = \Tr(X^3) + 3 Z^j_a X^a_b \tZ^b_j.}
The superpotential breaks the global symmetry classically to
$U(N_f)\times U(N_f)\times U(\tnf)\times U(1)_R$, and the instantons
break one of the $U(1)$ factors, so that the quantum
global symmetry of this
theory is
\eqn\globsymn{SU(N_f)\times SU(N_f)\times SU(\tnf)\times U(1)_B\times
U(1)_Z\times U(1)_R}
where we choose $U(1)_B$ and $U(1)_Z$ to be two arbitrary $U(1)$ symmetries
orthogonal to the $U(1)_R$ symmetry.

The quantum numbers of all
fields under the local symmetry $SU(N_c)$ and under the
global symmetry \globsymn\
may be easily computed, and they are summarized in the following table:
\input tables
\thicksize = 0pt
\thinsize = 0pt
\vskip 1.truecm
\begintable
$Q^i_a$ & $N_c$ & ( & $N_f$ ,& $1$, & $1$, & ${1\over N_c}$,&
$-{{N_f-N_c}\over N_c}$, &
$1+\ot{{\tnf-2 N_c}\over
N_f}$& ) \nr
$\tQ^a_{\ti}$ & $\overline{N_c}$ & ( & $1$, & $\overline{N_f}$, & $1$, &
$-{1\over {N_c}}$, &
${{N_f-N_c}\over{N_c}}$, &
$1+\ot{{\tnf-2 N_c}\over N_f}$ &
) \nr
${Z^j_a}$ & $N_c$ & ( & $1$, & $1$, & $\tnf$, & ${1\over N_c}$,
& $-{{(N_c+\tnf)N_f}\over{N_c \tnf}}$, &
$\tt$ &
) \nr
${\tZ_j^a}$ & $\overline{N_c}$ &
( & $1$, & $1$, & $\overline{\tnf}$, & $-{1\over N_c}$, &
${{(N_c+\tnf)N_f}\over{N_c \tnf}}$, &
$\tt$ &
) \nr
$X^a_b$ & $(N_c^2-1)$ & ( & $1$, & $1$, & $1$, & $0$, & $0$, &
$\tt$ & ) \nr
$W_{\alpha}$ & $(N_c^2-1)$ & ( & $1$, & $1$, & $1$, & $0$, & $0$, & $1$ & ) \nr
\endtable
\noindent
where the $R$ charge always pertains to the lowest component of the superfield.

\subsec{The ``magnetic" theory}

The ``magnetic" theory has a field content similar to that
of the ``electric" theory,
for gauge group $SU(2N_f+\tnf-N_c)$, with additional gauge singlet fields.
There are $N_f$ flavors of
quarks $q_i^a$ and $\tq^{\ti}_a$, $\tnf$ flavors of quarks $z_j^a$
and $\tz^j_a$ and an adjoint field $Y$.
Note that the non--abelian
global representations of the ``magnetic" quarks are conjugate
to those of the ``electric" quarks.
In addition to these there are
gauge singlet fields, which will be identified with mesons of the ``electric"
theory. These are $M^i_{\ti}$, $N^i_{\ti}$, $M^j_{\ti}$ and $M^i_j$, where
$i$ is an index of the first $SU(N_f)$ factor of the global symmetry,
$\ti$ is an index of the second $SU(N_f)$ factor of the global symmetry,
and $j$ is an index of $SU(\tnf)$.
The global symmetry group is the same as \globsymn.
The classical superpotential of the ``magnetic"
theory is
\eqn\nwmag{W_{mag} = \Tr(Y^3) + 3 z_j^a Y_a^b \tz^j_b + M^i_{\ti} q_i^a
Y_a^b \tq^{\ti}_b + N^i_{\ti} q_i^a \tq^{\ti}_a + M^i_j \tz^j_a q_i^a +
M^j_{\ti} z_j^a \tq^{\ti}_a.}
The non--renormalizable term in \nwmag\ is analogous to the one in
\dualk, and can be relevant (as in \dualk), due to non--perturbative
effects, when the ``magnetic" theory is strongly coupled.
We will see
in section 5 how to deal with this term when analyzing perturbations and flat
directions of the theory.

The local and global symmetry charges of the fields may easily be computed.
Making a choice of the $U(1)$ symmetries that will fit with the choice we made
in the ``electric" theory, we find them to be :
\thicksize = 0pt
\thinsize = 0pt
\vskip 1.truecm
\begintable
$q_i^a$ & $2N_f+\tnf-N_c$ & ( & $\overline{N_f}$, & $1$, & $1$, &
${1\over{2N_f+\tnf-N_c}}$, &
$-1$, &
${{2N_c-N_f-\tnf}\over{3N_f}}$ &
) \nr
${\tq}_a^{\ti}$ & $\overline{2N_f+\tnf-N_c}$ & ( & $1$, & $N_f$, & $1$, &
$-{1\over{2N_f+\tnf-N_c}}$, &
$1$, &
${{2N_c-N_f-\tnf}\over{3N_f}}$ &
) \nr
$M^i_{\ti}$ & $1$ & ( & $N_f$, & $\overline{N_f}$, & $1$, & $0$, & $0$, &
${{6N_f+2\tnf-4 N_c}\over{3N_f}}$ & ) \nr
$N^i_{\ti}$ & $1$ & ( & $N_f$, & $\overline{N_f}$, & $1$, & $0$, & $0$, &
${{8N_f+2\tnf-4 N_c}\over{3N_f}}$ & ) \nr
$M^j_{\ti}$ & $1$ & ( & $1$, & $\overline{N_f}$, & $\tnf$, & $0$, &
$-{{N_f+\tnf}\over {\tnf}}$, &
${{5N_f+\tnf-2 N_c}\over{3N_f}}$ &
) \nr
$M_j^i$ & $1$ & ( & $N_f$, & $1$, & $\overline{\tnf}$, & $0$, &
${{N_f+\tnf}\over {\tnf}}$, &
${{5N_f+\tnf-2 N_c}\over{3N_f}}$ &
) \nr
$z_j^a$ & $2N_f+\tnf-N_c$ & ( & $1$, & $1$, & $\overline{\tnf}$, &
${1\over {2N_f+\tnf-N_c}}$, &
$N_f \over \tnf$, &
$\tt$ &
) \nr
$\tz^j_a$ & $\overline{2N_f+\tnf-N_c}$ &
( & $1$, & $1$, & $\tnf$, &
$-{1\over {2N_f+\tnf-N_c}}$, & $-{N_f \over \tnf}$, &
$\tt$ &
) \nr
$Y$ & $((2N_f+\tnf-N_c)^2-1)$ & ( & $1$, & $1$, & $1$, &
$0$, & $0$, & $\tt$ & ) \nr
$W_{\alpha}$ & $((2N_f+\tnf-N_c)^2-1)$ &
( & $1$, & $1$, & $1$, & $0$, & $0$, & $1$ & ). \nr
\endtable
\noindent

\subsec{Comparing the two theories}

The first things which one should check in verifying a duality symmetry
are the 't Hooft anomaly matching conditions and the identification of
the gauge--invariant operators in the chiral ring, which we will perform in
this section. In later sections we will compare the mass perturbations of
the two theories and several of their flat directions as well. One should
also check that by performing the duality transformation twice one returns
to the original theory. This works in our case in the same way as in
the cases
described in \duals\ and in \dualk.

The global anomalies in both theories may easily be calculated to be :
\eqn\thooftn{\eqalign{
SU(N_f)^3 & \qquad N_c d^{(3)}(N_f) \cr
%SU(\tnf)^3 & \qquad 0 \cr
SU(N_f)^2 U(1)_B & \qquad d^{(2)}(N_f) \cr
%SU(\tnf)^2 U(1)_B & \qquad 0 \cr
SU(N_f)^2 U(1)_R & \qquad
{{N_c(\tnf-2N_c)} \over {3N_f}} d^{(2)}(N_f) \cr
SU(\tnf)^2 U(1)_R & \qquad - \tt N_c d^{(2)}(\tnf) \cr
SU(N_f)^2 U(1)_Z & \qquad (N_c-N_f) d^{(2)}(N_f) \cr
%SU(\tnf)^2 U(1)_Z & \qquad 0 \cr
U(1)_R & \qquad - \tt (N_c^2+1) \cr
U(1)_R^3 & \qquad
{1\over 27}(26 N_c^2 - 2 \tnf N_c - 26) + {{2N_c} \over {27 N_f^2}}
(\tnf - 2 N_c)^3 \cr
U(1)_B^2 U(1)_R & \qquad -{4\over 3} \cr
U(1)_Z^2 U(1)_R & \qquad
-{4\over 3} N_c^2 - {2\over {3\tnf}} N_c N_f^2 + \tt N_c \tnf -
{8\over 3} N_f^2 - {4\over 3} N_f \tnf + {8\over 3} N_f N_c \cr
}}
with all other anomalies vanishing trivially. Hence,
the anomaly matching conditions indeed hold.

Next, let us compare the gauge--invariant operators, starting with the
meson--like operators in the two theories. The ``normal" mesons of the
``electric" theory, which include at least one of
$Q$ or $\tQ$, may all be
identified with gauge singlets of the ``magnetic" theory :
\eqn\compm{M^i_{\ti} \sim Q^i_a \tQ_{\ti}^a; \qquad
M^j_{\ti} \sim Z^j_a \tQ_{\ti}^a ; \qquad M^i_j \sim Q^i_a \tZ_j^a. }
The mesons made out of a pair of $Z$-quark fields reside in the adjoint
and singlet representations
of the $SU(\tnf)$ flavor group. Naively, all of them may be
identified with the corresponding
mesons in the ``magnetic" theory, $Z^j \tZ_k \sim
z_k \tz^j$, but we shall see that this is only true for the adjoint
mesons and not for the singlet. The other ``normal" mesons of the ``magnetic"
theory are not in the chiral ring due to the equations of motion
associated with the superpotential \nwmag.

There are also generalized mesons which include
the adjoint field. In the ``electric" theory, these are the operators
$Q^i_a X^a_b \tQ^b_{\ti}$, which may be identified with $N^i_{\ti}$ in the
``magnetic" theory. All other generalized mesons, in both theories,
which include $Z$-quarks
or more than one power of the adjoint field, are not in the chiral ring.
This can be verified by using the equations
of motion associated with the superpotentials of the two theories,
which allow us to set them
to zero or to combinations of other singlet fields.

There is one more relatively simple operator in each of
the chiral rings of these
theories. It is $\Tr(X^2)$ in the ``electric" theory, and $\Tr(Y^2)$ in
the ``magnetic" theory. Superficially these two operators should be
identified, but in fact there is another operator with the same quantum
numbers, which is the singlet meson $Z^j_a \tZ_j^a$ in the ``electric"
theory and $z_j^a \tz^j_a$ in the ``magnetic" theory. Thus, there can
be a mixing between the two operators, and the most general possible
identification is of the form
\eqn\trident{\eqalign{
\Tr(Y^2) &\sim a \Tr(X^2) + b Z^j_a \tZ_j^a \cr
z_j^a \tz^j_a &\sim c \Tr(X^2) + d Z^j_a \tZ_j^a. \cr }}
We shall see that all the coefficients $a,b,c$ and $d$ are in fact non--zero.

The equations of motion resulting from the superpotential also
force the baryon--like operators in the chiral ring of the two theories to
be of the form :
\eqn\baryon{\eqalign{B^{[i_1,\cdots,i_k][i_{k+1},\cdots,i_n][j_1,\cdots,j_m]} =
\epsilon^{\alpha_1,\cdots,\alpha_{N_c}} & X^{\beta_1}_{\alpha_1} \cdots
X^{\beta_k}_{\alpha_k} Q^{i_1}_{\beta_1} \cdots Q^{i_k}_{\beta_k}
Q^{i_{k+1}}_{\alpha_{k+1}} \cdots Q^{i_n}_{\alpha_n} \cdot \cr &
Z^{j_1}_{\alpha_{n+1}}
\cdots Z^{j_m}_{\alpha_{n+m}} \cr }}
with $n+m=N_c$. It can easily be checked that these baryons in the ``electric"
theory have exactly the same global symmetry numbers as the baryons in
the ``magnetic" theory (constructed in the same way from $Y$'s, $q$'s and
$z$'s)
with ${\tilde k}=N_f-n+k$, ${\tilde n}=2N_f-n$ and $\tilde m=\tnf-m$.
Hence, all these operators may be identified.

More general hadronic operators involving quarks and anti--quarks also
appear in these theories \dualk.
These are, in the ``electric" theory, of the form
\eqn\caryon{\eqalign{
C^{[i_1,\cdots,i_k][j_1,\cdots,j_l]}_{[\ti_1,\cdots,\ti_m]
[k_1,\cdots,k_n]} = \epsilon^{\alpha_1,\cdots,\alpha_{N_c}}
\epsilon_{\beta_1,\cdots,\beta_{N_c}} & Q^{i_1}_{\alpha_1} \cdots
Q^{i_k}_{\alpha_k} Z^{j_1}_{\alpha_{k+1}} \cdots Z^{j_l}_{\alpha_{k+l}}
\tQ^{\beta_1}_{\ti_1} \cdots \tQ^{\beta_m}_{\ti_m} \cdot
\cr & \tZ^{\beta_{m+1}}_{k_1}
\cdots \tZ^{\beta_{m+n}}_{k_n} X^{\beta_{m+n+1}}_{\alpha_{k+l+1}} \cdots
X^{\beta_{N_c}}_{\alpha_{N_c}} \cr }}
with $k+l=m+n$. Similar operators exist in the ``magnetic" theory.
It can easily be checked that these operators may also be identified
between the two theories, with $\tilde k = N_f - k$, $\tilde m = N_f - m$,
$\tilde l = n$ and $\tilde n = l$.

The operators we analyzed in this section seem to be the
only independent gauge--invariant
operators in the chiral ring of the two theories (actually not all of the $B$
and $C$ hadrons are independent operators in the chiral ring),
and we have established that they
are identical in the two theories.

\subsec{Special cases of the duality}

As described above, the duality holds for all values of
$N_f$, $N_c$ and $\tnf$, as long as $2N_f+\tnf > N_c$.
The only difference for $\tnf=0$ or $N_f=0$ is
that some of the $U(1)$ symmetries disappear.
In the case of $\tnf=0$ one $U(1)$ symmetry disappears,
and the duality goes over to Kutasov's
duality \dualk. Hence,
the evidence we give in the next sections
for the duality may
be used as further support for Kutasov's duality.
Of course,
the quantum corrections may destroy the duality in some of these cases,
if they remove the origin of moduli space from the quantum moduli space.
This happens for instance in SQCD when $N_f\leq N_c$ \sei. However, we
see no a--priori
reason for this to occur in our theories, and in particular we have a
solution for the 't Hooft anomaly conditions whenever $2N_f+\tnf>N_c$.

For $N_f=0$ two of the $U(1)$ symmetries disappear (for general $\tnf$),
and in particular we no longer have a $U(1)_R$ symmetry.
In both theories we are left
with only the adjoint field and with the $Z$-quarks,
which have an $R$-charge of $2/3$ and an
interaction with the adjoint field.
The duality in this case takes the gauge group $SU(N_c)$ to $SU(\tnf-N_c)$,
like the duality described by Seiberg \duals\ without the adjoint field.
The ``electric" theory is now asymptotically free for $\tnf < 2N_c$, and the
``magnetic" theory is asymptotically free for $\tnf > 2N_c$. Therefore, we do
not have an IR fixed point for any value of $\tnf$. This is consistent
with the fact that we have no $U(1)_R$ symmetry in this case.
For $\tnf=2N_c$ the $U(1)_R$ symmetry remains unbroken, and we find that
the original theory is equal to the dual theory (except for the flavor
representations of the quarks). In this case the theory
may be exactly conformal, and the duality we analyze
may be related to the $N=2$
duality discovered by Seiberg and Witten \refs{\swi-\klyt}. Note that the
coupling of the quarks to the adjoint fields is the same as in the $N=2$
theory. The main difference seems to be the existence of the $\Tr(X^3)$ term
in the superpotential, which explicitly breaks the $N=2$ symmetry (and the
$R$ charge associated with it). It is possible that by turning on a
perturbation proportional to this operator in the $N=2$ theory, one may
flow from the $N=2$ duality (with $N_f=2N_c$ and $N_c>2$)
to our duality for $N_f=0$ and $\tnf=2N_c$. From there, one can flow
to other values of $\tnf$ as described in the next sections. Unfortunately,
we have not identified any flows which increase the value of $N_f$.
We, therefore, still do not know how to
flow to all cases from this one. However, one can flow from the
case of general $N_f$ to the case of $N_f=0$. Hence, the understanding
of the case
of $N_f=0$ may help in understanding the general case as well.

\newsec{Mass perturbations of the dual theories}

Another comparison which one can make between
the two theories concerns their behavior under mass perturbations.
In this section we will analyze mass perturbations of the dual theories.
We will ignore the possible quantum corrections to the superpotential
whenever this is possible. In most cases we will see that this indeed
gives a flow consistent with the duality. However, it seems clear that, as
in Seiberg's duality \duals, whenever one of the gauge groups is
completely broken the quantum corrections become important. Therefore,
all the flows we describe here, and in the next section when we analyze
the flow along flat directions, are presumably relevant only when neither
gauge group is completely broken.

\subsec{Quark mass perturbations of the dual theories}

When we add a mass to an ``electric" $Q$-quark, i.e. a term $m M^{N_f}_{N_f}$
in the superpotential, the behavior is the same as in Kutasov's original theory
\dualk. In
the ``electric" theory the quarks $Q^{N_f}$ and $\tQ_{N_f}$ become massive,
reducing $N_f$ by one. In the ``magnetic" theory the equation of motion of
$M^{N_f}_{N_f}$ forces us to give VEVs to $q_{N_f}$, $\tq^{N_f}$ and $Y$,
reducing $N_f$ by one and the number of colors by two.
The resulting theories are again dual under the duality transformation.

Adding a mass to the $Z$-quarks is more subtle, due to the identification
\trident, and will be discussed later.
However, it is relatively simple to perturb the
theory by a mass operator of the form $m Q^{N_f} \tZ_{\tnf} \sim m
M^{N_f}_{\tnf}$. In the ``electric" theory, this gives a mass to $Q^{N_f}$
and to $\tZ_{\tnf}$, and leaves $Z^{\tnf}$ without a coupling to the adjoint
field, i.e. it becomes a regular $Q$-quark.
Thus, we reduce $\tnf$ by one, leaving $N_c$ and $N_f$
as they were. In the ``magnetic" theory, the equation of motion of
$M^{N_f}_{\tnf}$ forces $\tz^{\tnf}$ and $q_{N_f}$ to get VEVs, breaking
the ``magnetic" gauge symmetry to $SU(2N_f+\tnf-N_c-1)$. Let us choose
the components that get non--zero VEVs to be $q^1_{N_f}$ and $\tz^{\tnf}_1$.
Then, $\tz^{\tnf}$ and $q_{N_f}$ are swallowed by the Higgs mechanism, and
the fields $z_{\tnf}^a$ and $Y^1_a$ get a mass from the superpotential
(as well as some other fields). When integrating out the massive fields
we find that $Y^a_1$ now couples just like
a regular $q$-quark. Hence,
$N_f$ does not change while $\tnf$ is reduced by one.
This is exactly the dual
of the result we found in the ``electric" theory. Thus, this perturbation
preserves the duality.

\subsec{Mass perturbations of the $Z$-quarks}

Since the identification of the singlet $Z$-meson in \trident\ is complicated,
let us analyze first the
$Z$ mass terms which are in the adjoint
representation of the flavor group.
An example of such a term is
$m(Z^1 \tZ_1 - Z^2 \tZ_2) \sim m(z_1 \tz^1 - z_2 \tz^2)$. Superficially
this gives a mass to two $Z$-quarks in both theories, reducing $\tnf$ by two
in both theories. This is obviously inconsistent with the duality. However,
because of the coupling of the $Z$-quarks to the adjoint field, we must
be careful in interpreting $m$ as the actual ``physical" mass of the quarks.
There is a vacuum in which $X$ ($Y$) does not get a VEV, and then $m$ is indeed
the mass of the quarks. However, there could also be some other vacua.
There is a-priori no reason
to identify the vacuum in which the VEVs of all fields are zero
in the ``electric" theory with the same vacuum of the ``magnetic" theory.
The only demand is to identify all the gauge--invariant operators in the
corresponding theories. Let us try, therefore, to find
a vacuum for which $X$ gets a VEV in the ``electric" theory,
in the theory with finite $m$ (the analysis
of the ``magnetic" theory is analogous). Note
that this theory does not have an $R$ symmetry, and, therefore,
in the IR $m$ must
flow either to zero or to infinity.
One can easily find that the following VEVs
satisfy all the $F$-term and $D$-term constraints of the perturbed theory :
\eqn\vevszz{\langle X^1_1 \rangle = -\ot m; \langle X^2_2 \rangle = \ot m;
\langle Z^1_1 \rangle = \ot m; \langle \tZ^1_1 \rangle = -\ot m; \langle
Z^2_2 \rangle = \ot m; \langle \tZ^2_2 \rangle = -\ot m}
with all other VEVs vanishing. Along this flat direction we find that the
gauge group is broken to $SU(N_c-2)$, and all components of $Z^1,\tZ_1,
Z^2$ and $\tZ_2$ become massive, either by contributions
from the superpotential or through
the Higgs mechanism. Hence, along this flat direction we flow to a theory
with both $\tnf$ and the number of colors reduced by two. This is dual to the
theory we found for zero VEVs, which had $\tnf$ reduced by two without a
change in the number of colors. Thus, when we include all flat directions,
this mass perturbation preserves the duality symmetry.

Let us now discuss the perturbation by the mass operator of one of
the $z$-quarks in the ``magnetic" theory. In the ``magnetic" theory,
this perturbation is $m z_1 \tz^1$ which we may write as
\eqn\mpert{m((\tnf-1) z_1 \tz^1 - z_2 \tz^2 - \cdots - z_{\tnf} \tz^{\tnf}
+ z_j \tz^j) / \tnf}
dividing it into components which are in the adjoint and singlet
representations of the flavor group $SU(\tnf)$. Now, according to
\trident, we should identify this operator in the ``electric" theory
with
\eqn\epert{m((\tnf-1) Z^1 \tZ_1 - Z^2 \tZ_2 - \cdots - Z^{\tnf} \tZ_{\tnf}
+ c \Tr(X^2) + d Z^j \tZ_j) / \tnf}
which equals
\eqn\nepert{m((\tnf+d-1) Z^1 \tZ_1 + (d-1) (Z^2 \tZ_2 + \cdots + Z^{\tnf}
\tZ_{\tnf}) + c \Tr(X^2)) / \tnf.}

As in the previous section,
it is not obvious that $m$ is actually the ``physical" mass of
$z_1$ and $\tz^1$, because of the $z_1 Y \tz^1$ coupling in the
superpotential. However, there is certainly a flat direction of
the ``magnetic" theory for which the VEV of $Y$ vanishes, and then $m$
is indeed the ``physical" mass of these fields. Along this flat direction,
we find in the ``magnetic" theory that $\tnf$ decreases by one with no
change in $N_f$ and in the gauge group. Thus, we should be able to find
in the ``electric" theory the dual of this result, i.e. a flat direction
along which $N_f$ is unchanged, and both $\tnf$ and $N_c$ decrease by one.
By demanding the existence of such a flat direction
we will be able to fix the values of $c$ and $d$ in equation \trident.

The most general VEVs we can have in the ``electric" theory which
will satisfy the $D$-term equations, break the color group to $SU(N_c-1)$,
break the $Z$-flavor group to $SU(\tnf-1)$ and will not affect the $Q$-flavor
group, are of the form (up to global and local transformations)
\eqn\evevs{\eqalign{
\langle X^1_1 \rangle &= A \cr
\langle X^j_j \rangle &= B \qquad (j=2,\cdots,N_c) \cr
\langle Z^1_1 \rangle &= \langle \tZ^1_1 \rangle = F \cr }}
with some constants $A, B$ and $F$. All other VEVs vanish.
We can obtain several equations
relating $A, B, F, c$ and $d$. First of all, since $X$ must be traceless,
we find that
\eqn\eqna{A+(N_c-1)B=0.}
Next, we demand that these VEVs be solutions
of the equations of motion resulting from the perturbed superpotential.
The equation of motion of $X$ in the perturbed theory implies that the
matrix $3X^2+{{2mc}\over \tnf}X + 3 Z^j \tZ_j$ must be proportional to the
identity matrix. This gives an additional equation,
\eqn\eqnb{3B^2+{{2mc}\over \tnf}B = 3A^2 + {{2mc}\over \tnf}A + 3 F^2.}
The equations of motion of $Z^1$ and $\tZ_1$ lead to another equation,
\eqn\eqnc{3A+m{{\tnf+d-1}\over \tnf} = 0.}
We need two more equations to determine all the constants, and these
will be obtained by demanding that all the fields $Z^j$, $\tZ_j$ for
$j=2,\cdots,\tnf$ and $X^j_k$
for $j,k=2,\cdots,N_c$ remain massless along this flat direction.
(Recall that we want to flow to the ``electric" theory with $\tnf-1$ massless
flavors of
$Z$-quarks and $N_c-1$ colors). By demanding that $\tnf-1$ flavors of
$Z$-quarks remain
massless we find
\eqn\eqnd{3B+m{{d-1}\over \tnf}=0}
and by demanding that the $SU(N_c-1)$ adjoint components of $X$ remain
massless we find
\eqn\eqne{3B+{{cm}\over \tnf} = 0.}
The above five equations have a unique solution for $A,B,F,c$ and $d$.
Thus, we have found the values of $c$ and $d$ in \trident :
\eqn\solcd{
c=-{\tnf \over N_c}; \qquad d=1-{\tnf \over N_c}.}
We have
also found, in the ``electric" theory, a flat direction along which
the theory flows to the dual of the theory that we have found when we
considered the ``magnetic"
theory. The duality is, therefore, consistent with this mass perturbation.

We can find the other constants in equation \trident\ by using
the inverse equation
\eqn\ntrident{\eqalign{
\Tr(X^2) &\sim \ta \Tr(Y^2) + \tb z_j^a \tz^j_a \cr
Z^j_a \tZ_j^a &\sim \tc \Tr(Y^2) + \td z_j^a \tz^j_a. \cr }}
By repeating the previous analysis with a perturbation proportional
to $Z^1 \tZ_1$, we can easily find that
\eqn\solcdn{\tc=-{\tnf
\over{2N_f+\tnf-N_c}}; \qquad \td=1-{\tnf \over {2N_f+\tnf-N_c}}.}
The
consistency of \trident\ and \ntrident\ requires that $c\ta + d\tc =
a\tc + c\td = b\ta + d\tb =
a\tb + b\td = 0$ and that $c\tb + d\td = b\tc + d\td = a\ta + b\tc =
a\ta + c\tb = 1$. The solution to all these equations is
\eqn\solab{a={{N_c-2N_f}\over
N_c};\qquad b=-{{2N_f}\over N_c};\qquad \ta={{\tnf-N_c}\over {2N_f+\tnf-N_c}};
\qquad
\tb=-{{2N_f}\over{2N_f+\tnf-N_c}}.}
The identification of the operators with
these values of the constants is the only one which is
consistent with the duality, and we will assume it to hold.

\subsec{Mass perturbations of the adjoint fields}

Next, we consider the behavior of the theories under a mass perturbation
associated with a mass operator for the adjoint fields.
We shall start with the case $\tnf=0$, and later
 analyze the differences which arise in the case $\tnf > 0$. When $\tnf=0$, the
identification of the mass operator for the adjoint fields is
necessarily $m \Tr(X^2) \propto m\Tr(Y^2)$. As in the previous discussions,
we ignore the quantum corrections to the superpotential. However,
for this perturbation we will find that (at least) in some cases
they do play an important role.

When we add this operator to the ``electric" theory,
the equation of motion of
$X$ implies that $3X^2+2mX$ (where $X$ here is the VEV of the field $X$)
must be proportional to the identity matrix. One
trivial solution of this equation of motion
is $X=0$. In this case we will have no breaking
of the gauge symmetry, and $m$ will be the ``physical" mass of the field
$X$. However, there are also other solutions. Since $X$ obeys $\Tr(X)=0$,
it can easily be seen that the most general solution (up to gauge
transformations) for which no other fields get VEVs is $X^i_i=A,i=1,\cdots,
k$ and $X^j_j=B,j=k+1,\cdots,N_c$, where $0<k<{N_c/2}$,
$A=\tt m {{N_c-k}\over{2k-N_c}}$
and $B=\tt m {k\over {N_c-2k}}$.
The previous
case of $X=0$ may be obtained from this one by taking $k=0$. To analyze
the spectrum along this flat direction, let us write the matrix $X$ in
the form
\eqn\xform{X=\pmatrix{ AI+X_1 & X_2 \cr X_3 & BI+X_4 \cr }}
where $I$ is the identity matrix and we divide the rows/columns of $X$
into groups of size $(k,N_c-k)$. By examining the superpotential, we can
see that the fields corresponding to the
matrices $X_1$ and $X_4$ become massive, while those corresponding to $X_2$
and $X_3$ remain massless and are swallowed by the Higgs mechanism.
The gauge group breaks into $SU(k)\times
SU(N_c-k)\times U(1)$.
The quarks and anti--quarks are
divided into two groups. The first
one includes those which are charged under $SU(k)$ and singlets
of $SU(N_c-k)$, while the other includes those
which are charged under $SU(N_c-k)$ and singlets
of $SU(k)$.
When we integrate out the massive fields, the generated superpotential
of the ``electric" theory along this flat
direction turns out to be zero.
Thus, we flow to a theory containing two copies of
supersymmetric QCD, one with $N_f$ flavors and gauge group
$SU(k)$, and the other with $N_f$ flavors and gauge group $SU(N_c-k)$. There is
an additional local $U(1)$ symmetry whose charge is proportional
to the baryon number in
each of the two theories.

Let us now analyze the same perturbation in the ``magnetic" theory. The
analysis of the possible VEVs of $Y$ is similar to the one we have
performed in
the ``electric" theory. The possible breakings of the gauge group are to
$SU(\tk)\times SU(2N_f-N_c-\tk)\times U(1)$ with $0 \leq \tk < N_f-N_c/2$.
The only difference is that in the ``magnetic" theory there is also a coupling
of the form $M q Y \tq$, and we must understand what a large VEV for $Y$ does
to this non--renormalizable term. We postpone the analysis of this issue
to section 5.2
and mention here only the results. We find that a large
VEV for $Y$ causes, after integrating out the massive fields, this term
to go over to a term of the form $M^i_{\ti} q_i^a \tq^{\ti}_a$,
plus non--renormalizable terms
which appear to be irrelevant. The constant in front of this term depends
on the VEV of $Y$, and thus we will find a different constant for the fields
$q^a \tq_a$ with $a=1,\cdots,\tk$ and for the fields $q^a \tq_a$ with $a=\tk+1,
\cdots,2N_f-N_c$. We can then add this term to the original $N q \tq$ term in
the superpotential. We find that we have two independent meson fields in the
$(N_f,\overline{N_f})$ representation of the $SU(N_f)\times SU(N_f)$ flavor
group. One of the mesons (a linear combination of $M$ and $N$)
couples only to the first $\tk$ colors of quarks while
the other couples only to the last $2N_f-N_c-\tk$ colors of quarks. As in the
``electric" theory, the fields $Y_2$ and $Y_3$ also remain massless, and
are swallowed by the Higgs mechanism.
We thus flow to two copies of Seiberg's ``magnetic" theory \duals,
one with $N_f$ flavors and a gauge group $SU(\tk)$ and the other with
$N_f$ flavors and a gauge group $SU(2N_f-N_c-\tk)$. As in the ``electric"
theory, there is also a local $U(1)$ whose charge is proportional to the
baryon number in each of the two theories.

Now that we have found the resulting theories in both cases, we should see how
they are dual to each other. This means that we should match the flat
directions which we have found, along which (when adding the mass
perturbation) one can flow, between the two theories.
Let us start with the case of
$N_f \geq N_c$ and $0<k<N_c/2$.
This condition means that in the ``electric" theory we have
$N_f > k$ and $N_f > N_c-k$. From Seiberg's
results \duals\ we know that the theories of $N_f$ flavors and $k$ colors
and of $N_f$ flavors and $N_c-k$ colors each have a dual IR description.
This description includes
$N_f$ flavors, $N_f-k$ colors and additional meson--like
singlet fields for the first
theory, and $N_f$ flavors, $N_f-N_c+k$ colors and additional meson--like
singlet
fields for the second theory. (Strictly speaking, Seiberg's results
regarding duality apply
only for $N_f > N_c+1$, but we can use them also for $N_f=N_c+1$ since then
they lead to a dual description in terms of mesons and baryons with no
gauge symmetry \sei). This is exactly the theory we find when going along
the flat direction of $\tk=N_f-N_c+k$ in the ``magnetic" theory. Thus,
along these flat directions we find that our duality goes over to two copies
of Seiberg's duality \duals,
with an additional local $U(1)$ symmetry coupling to
the baryon number of the respective ``electric" and
``magnetic" theories.
The same analysis applies to the case of $N_f \leq N_c$ and
$0<\tk<N_f-N_c/2$ when we
interchange the roles of the ``electric" and ``magnetic" theories.

We still need to understand the cases of $k=0$ and $\tk=0$, the cases
of $0 < \tk < N_f-N_c$ when $N_f > N_c$ and the cases of $0 < k < N_c-N_f$
when $N_f < N_c$, for which we did not find dual theories
in the discussion above.
These cases all lead, at least in one of the theories (when identifying
$\tk=N_f-N_c+k$), to a SQCD theory
in which the number of quark flavors is smaller
or equal to the number of colors. For these theories it is known that
the quantum effects are important, and the quantum moduli space is not
equal to the classical moduli space \sei. We will
assume that the quantum effects along these flat directions in our
theories are the same as
in the corresponding SQCD theories \sei.
We shall see that this assumption
is consistent with the duality.

Let us start with the case of $k=0$ for $N_f > N_c$.
According to the above discussion, this
should be dual to the case of $\tk=N_f-N_c$, which gives
rise to a gauge theory of
$SU(N_f)\times SU(N_f-N_c)\times U(1)$ in the ``magnetic" theory. In the
``electric" theory we get just SQCD with $N_f$ flavors and $N_c$ colors,
for which we have \duals\ a dual description
in terms of a ``magnetic" theory based on the gauge group $SU(N_f-N_c)$.
Thus,
we need to understand the ``extra" $SU(N_f)$ gauge theory with $N_f$ flavors
which we obtain in the
``magnetic" theory. According to \sei, when performing
an exact quantum analysis of this theory (without
the additional singlets which we have here), it has an IR description in
terms of mesons $m_i^{\ti}$ and baryons $B$ and $\tB$, constrained by
$\det(m)-B \tB = \Lambda^{2N_f}$. With the additional singlets we,
therefore, find
that our $SU(N_f)$ gauge theory is equivalent to the theory
of mesons and baryons with this constraint and with an
additional superpotential of the form $M^i_{\ti} m_i^{\ti}$ ($M$ denotes
the mesons which couple, in the ``magnetic" theory,
to the quarks which are charged under the local $SU(N_f)$, as
described above). Integrating
out the field $M$ we find that $m=0$. Hence, we are left just with
the fields $B$ and $\tB$ constrained by $B \tB = -\Lambda^{2N_f}$.
These fields are charged under the local $U(1)$ symmetry, with opposite
charges.
We claim that this theory, when combined with the $SU(N_f-N_c)$
``magnetic" theory and with a $U(1)$ gauge symmetry coupling the two,
is equivalent to the ``electric" $SU(N_c)$ gauge theory.
To form a gauge--invariant operator in the ``magnetic" theory, we must
take an operator in the $SU(N_f-N_c)$ ``magnetic" theory (these are known
to be in a one-to-one correspondence with the operators of the ``electric"
theory \duals) and add to it an appropriate power of $B$ or $\tB$ so that
the result is invariant under the local $U(1)$ symmetry. Obviously there
is just one way to do this. Thus, the resulting theories along
these flat directions are also dual.
A similar analysis holds for the
case $\tk=0$ when $N_f < N_c$.

The other cases we had problems with were those of $k<N_c-N_f$ when
$N_f < N_c$, and those of $\tk<N_f-N_c$ when $N_f > N_c$. In all of these
cases one of the gauge theories we find along this flat direction has
a number of flavors smaller than the number of colors. In this case it
is known that SQCD does not have a stable vacuum, due to quantum effects.
Thus, it is possible that these flat directions may not even exist
in the full quantum theory, due to the quantum effects that we have ignored.
In any case we expect the quantum corrections to be important along these
flat directions and we shall not analyze them here any further.
Except for these cases, all flat directions in the ``electric" theory are
identified with flat directions in the corresponding ``magnetic" theory.
In all cases we
flow to theories which are already known to be dual, either by the
description of SQCD for $N_f=N_c$ and $N_f=N_c+1$
in terms of bound states \sei, or by
the duality transformation of SQCD \duals.

The case of $\tnf>0$ is similar to the previous case. However, according
to \trident, in this case the mass operator
$m\Tr(X^2)$ is identified with a non--trivial
linear combination of $\Tr(Y^2)$ and $z_j \tz^j$ :
\eqn\ntrident{m\Tr(X^2) \sim m(\ta \Tr(Y^2) + \tb z_j \tz^j).}
In the ``electric" theory, the analysis of the VEVs of $X$
is the same as in the previous case
(assuming that the $Z$-quarks do not get VEVs). The VEV
of $X$ gives a mass to all $Z$-quarks (except in the case $k=0$), and,
therefore,
we flow to the same theories we discussed above (for $k>0$). In the ``magnetic"
theory, we must now find VEVs for $Y$ and for the $z$-quarks which will solve
the equations of motion coming from
the superpotential perturbed by the terms $\Tr(Y^2)$
and $z_j \tz^j$. The solutions for which the $z$-quarks do not get VEVs
are dual to solutions in which the $Z$-quarks do get VEVs. Hence, we will
analyze here the case in which all $z$-quarks get VEVs.
These VEVs must be orthogonal and of equal magnitude for all the $z$-quarks
and anti--quarks (i.e. they have the form $z_j^a=A \delta_j^a, \tz^j_a =
A\delta^j_a$).
In this case
we find that we flow exactly to the theories described above. The VEVs
of $z$, $\tz$ and $Y$ break the gauge
symmetry to $SU(\tk)\times SU(2N_f-N_c-\tk)\times U(1)$. All the $z$-quarks
and components of $Y$ either become massive or
are swallowed by the Higgs mechanism.
The resulting
theories are exactly the theories discussed
above, for which we already investigated the duality.
The cases in which the $Z$-quarks get VEVs (and the $z$-quarks do not), and
the cases
in which we add a term $\Tr(Y^2)$ instead of $\Tr(X^2)$ are all analogous.
Note that we assumed here that $N_f>N_c/2$. If this does not hold, we
expect the quantum corrections always to be important, since we always
seem to flow to a SQCD theory with a number of flavors smaller
than the number of colors.

We must still discuss what happens when $k=0$ and $\tnf>0$. In this case
in the ``electric" theory the $Z$-quarks do not become massive, and
integrating out the field $X$
generates a (non--renormalizable)
interaction between them of the form
$W=Z^j_a \tZ_k^a Z^k_b \tZ_j^b$, which forces their $R$ charge to be ${1
\over 2}$.
Thus, we flow to a SQCD theory with $N_f$ flavors of one type, $\tnf$
flavors of another type (with $R={1\over 2}$)
and
$N_c$ colors. In the ``magnetic" theory, along the corresponding flat
direction,
only $Y$ gets a VEV, which breaks the gauge group to $SU(N_f)\times
SU(N_f+\tnf-N_c)\times U(1)$. The mass of the $z$-quarks gets contributions
both from the explicit mass term \ntrident\ and from the VEV of $Y$ through
the superpotential. For the duality to hold, these terms must cancel out
for the $z$-quarks which are
charged under the $SU(N_f+\tnf-N_c)$ gauge group. Using
the values of $\ta$ and $\tb$ which we found above we see that this indeed
occurs.
The ``magnetic" theory turns out to be a ``sum" of two theories.
The first is a
``magnetic" theory of $N_f$ flavors and
$N_f$ colors (which behaves as in the discussion above). The second is
a theory of $N_f$ ``normal" flavors and $\tnf$ additional flavors,
which have a superpotential analogous to the one we
have found in the ``electric" theory,
with gauge group $SU(N_f+\tnf-N_c)$.
We claim that this theory is dual to the theory which we have found when
we considered the corresponding flow
from the
``electric" theory. It includes couplings of the form $M q \tq + M q \tz
+ M z \tq$, but not $M z \tz$, where the mesons may be identified with the
operators $Q \tQ$, $Z \tQ$ and $Q \tZ$ in the ``electric" theory.
This duality satisfies the usual requirements, i.e.
the 't Hooft anomaly conditions for the global symmetry
group (which is the same as the original global symmetry group) and the
identification of the bound states. In fact, we can flow to this duality
also directly from Seiberg's duality \duals, starting from a SQCD theory with
$N_f+\tnf$ flavors and $N_c$ colors and perturbing it by an
operator of the form $M^j_k M^k_j$, where we include all mesons composed of
the last $\tnf$ flavors.
Thus, we find that also this case is
compatible with the duality.

\newsec{Flat directions in the dual theories}

In this section we will analyze some of the flat directions of the
dual theories, verifying that their behavior is consistent with
the duality. Since we do not analyze here the quantum corrections to the
superpotentials of the two theories, we cannot prove that these flat
directions indeed exist and behave as we describe.
However, all the terms one can imagine in the
superpotential do not seem to eliminate these flat directions of the
classical theory. We shall, therefore, assume their existence in the full
theory. The
consistency of our results strongly suggests that these flat directions
do indeed exist. In this section we study, for simplicity,
only flat directions along which just one gauge--invariant operator
gets a VEV while all others do not.
We find that along some flat directions the theory flows to another
dual theory, along some other flat directions
the ``electric" and ``magnetic" theories both flow to the same theory in
the IR, and along one flat direction the duality flows
to Seiberg's duality \duals.
The discussion of the flat direction
for which the meson
operator $M^i_{\ti}$ gets a VEV is more complicated and, therefore,
is postponed to the next section.

\subsec{Mesonic flat directions}

In Seiberg's duality \duals, the effect of giving a meson a VEV on the
``electric" (``magnetic") theory is equivalent to the effect of adding
a mass term (proportional to the meson operator) on the ``magnetic"
(``electric") theory. For the dual theories we analyze, it can easily be seen
that giving the meson $M^i_{\ti}$ a VEV
is similar to adding a term proportional
to $N^i_{\ti}$ to the superpotential. Similarly,
giving the generalized meson $N^i_{\ti}$ a VEV is dual in this sense
to adding a perturbation proportional to $M^i_{\ti}$ (i.e. a mass term in the
``electric" theory). Thus, the analysis of the flat direction along which
only $N^i_{\ti}$ gets a VEV follows the
analysis of the $Q$-quark mass perturbation described in the previous
section.

Let us give a VEV only to the generalized meson
$Q^1 X \tQ_1 \sim N^1_1$, keeping the VEVs of
all other gauge--invariant operators zero. In the ``electric"
theory, this corresponds (up to color transformations) to
$Q^1_1$, $\tQ^2_1$ and $X^1_2$ getting VEVs, which can be checked to be
a flat direction of the theory. The color group breaks to $SU(N_c-2)$,
with $N_f-1$ flavors remaining massless.
Some components of $X$ also get a mass from the
superpotential, leaving the adjoint of $SU(N_c-2)$ massless.
Thus,
along this flat direction we flow to the ``electric" theory with $N_f-1$
flavors and $N_c-2$ colors. $\tnf$ remains unchanged along this
flat direction.
In the ``magnetic" theory, along this flat direction only $N^1_1$ gets
a VEV. This leads to a mass term for $q_1$ and $\tq^1$, and we
flow to the ``magnetic"
theory with ($N_f-1$) $q$-flavors, $2N_f+\tnf-N_c$ colors, and with
$\tnf$ unchanged. Since
$2N_f+\tnf-N_c=2(N_f-1)+\tnf-(N_c-2)$ this is exactly the dual to the
``electric" theory
which we found
above. We conclude that the duality holds
along
this flat direction.

Another flat direction of these theories is realized by giving a VEV to
$Z^j \tQ_{\ti} \sim M^j_{\ti}$. The analysis of this flat direction
is similar to the analysis
of the perturbation by this operator carried out in the previous section.
We should just interchange the ``electric" and ``magnetic" theories.
In the ``electric" theory, $\tnf$ decreases by one and the gauge group
breaks to $SU(N_c-1)$, while in the ``magnetic" theory, $\tnf$ decreases
by one without a change in the gauge group. Again,
the duality is preserved.

Yet another flat direction is obtained by
giving a VEV to $Z^2 \tZ_1 \sim z_1 \tz^2$.
In both theories, this reduces the number of colors by one. The quarks
$Z^2$,$\tZ_1$,$z_1$ and $\tz^2$ are all swallowed by the Higgs mechanism,
while the quarks $Z^1$,$\tZ_2$,$z_2$ and $\tz^1$, as well as some components
of $X$ and $Y$, get masses from the superpotential. Thus, in both theories
$\tnf$ is reduced by two while the number of colors is reduced by one.
This result is compatible with the duality.
The discussion of the flat direction along
which a diagonal $Z$-meson gets
a VEV is analogous to the discussion of the perturbation by the $Z$-mass
operator in the previous section and, therefore, will not be
repeated here.

\subsec{The baryonic flat direction}

In Seiberg's dual theories \duals, the baryon composed of quarks is
equivalent to the baryon composed of dual quarks. In \dualo\
the flat direction along which this baryon gets a VEV was analyzed.
It was found that the
same IR theory results from both the ``electric" and the
``magnetic" theories. In the dual theories analyzed here
we also expect the gauge
groups to be completely broken along this flat direction, and to
find the same IR effective theory resulting from both theories.
However, in the present case the baryon which is
composed of only quarks in one theory
is equivalent to the baryon composed of quarks and adjoint fields in
the dual theory. We shall analyze here the simplest case, in
which the baryon made of only quarks (in one of the theories)
gets a VEV.
It seems quite obvious (and has been checked in several cases)
that the behavior along flat
directions for which other baryons (which are made in both theories
out of the corresponding $Q$ and $Z$ quarks and adjoint fields)
is similar. We shall find in this
case
that the ``electric"
and the ``magnetic" theories indeed give rise to the same IR theory.

As described in section 2, the baryon operator in the
``electric" theory is given by
\eqn\bare{\eqalign{
B_{el}^{[i_1,\cdots,i_k][i_{k+1},\cdots,i_n][j_1,\cdots,j_m]} =
\epsilon^{\alpha_1,\cdots,\alpha_{N_c}} & X^{\beta_1}_{\alpha_1} \cdots
X^{\beta_k}_{\alpha_k} Q^{i_1}_{\beta_1} \cdots Q^{i_k}_{\beta_k}
Q^{i_{k+1}}_{\alpha_{k+1}} \cdots Q^{i_n}_{\alpha_n} \cdot \cr &
Z^{j_1}_{\alpha_{n+1}}
\cdots Z^{j_m}_{\alpha_{n+m}} \cr }}
with $m+n=N_c$,
and in the ``magnetic" theory by
\eqn\barm{\eqalign{
B_{mag}^{[{\ti}_1,\cdots,{\ti}_{\tk}][\ti_{\tk+1},\cdots,\ti_{\tn}]
[\tj_1,\cdots,\tj_{\tm}]} =
\epsilon_{\alpha_1,\cdots,\alpha_{2N_f+\tnf-N_c}} &
Y_{\beta_1}^{\alpha_1} \cdots
Y_{\beta_{\tk}}^{\alpha_{\tk}} q_{\ti_1}^{\beta_1} \cdots
q_{\ti_{\tk}}^{\beta_{\tk}}
q_{\ti_{\tk+1}}^{\alpha_{\tk+1}} \cdots q_{\ti_{\tn}}^{\alpha_{\tn}}
\cdot \cr &
z_{\tj_1}^{\alpha_{\tn+1}}
\cdots z_{\tj_{\tm}}^{\alpha_{\tn+\tm}} \cr }}
with $\tm+\tn=2N_f+\tnf-N_c$.
The identification between the operators in the two theories is by
$\tk=N_f-n+k$, $\tn=2N_f-n$, $\tm=\tnf-m$,
and by $\epsilon^{i_1,\cdots,i_k,\ti_{\tk+1},\cdots,\ti_{\tn}}=
\epsilon^{\ti_1,\cdots,\ti_{\tk},i_{k+1},\cdots,i_n} =
\epsilon^{j_1,\cdots,j_m,\tj_1,\cdots,\tj_{\tm}} = 1$.

Let us choose the baryon
made of the first
$N_c$ $Q$-quarks, $B_{el}^{[\ ][1,\cdots,N_c][\ ]}$, to get a non--zero VEV.
This baryon only exists for $N_f \geq N_c$ which we will assume in this
subsection. In the ``magnetic" theory this baryon is equivalent to
$B_{mag}^{[N_c+1,\cdots,N_f][1,\cdots,N_f][1,\cdots,\tnf]}$.
In the ``electric" theory, the first $N_c$
$Q$-quarks get VEVs and are swallowed by the Higgs mechanism, with the
gauge group being completely broken (one chiral superfield remains massless,
labeling the flat direction. We shall ignore it since it is a
non--interacting singlet). All other fields remain massless.
As in Seiberg's theory \dualo, the global
symmetry breaks to
\eqn\globsymb{SU(N_c)\times SU(N_f-N_c) \times SU(N_f) \times
SU(\tnf) \times
U(1)_{\hat B} \times U(1)_{\hat Z} \times U(1)_{\hat R},}
and the charges of the massless
fields turn out to be :
\thicksize = 0pt
\thinsize = 0pt
\vskip 1.truecm
\begintable
$Q^{\hat i}_a$ & ( & $N_c$, & $N_f-N_c$ ,& $1$, & $1$, &
${N_f\over {(N_f-N_c)N_c}}$,&
$-{N_f\over N_c}$, &
${{3N_f+\tnf-2 N_c}\over
{3(N_f-N_c)}}$&
)  \nr
$\tQ^a_{\ti}$ & ( & $\overline{N_c}$, & $1$, & $\overline{N_f}$, & $1$, &
$-{1 \over {N_c}}$, &
${{N_f-N_c}\over{N_c}}$, &
$1+\ot{{\tnf-2 N_c}\over N_f}$ &
) \nr
${Z^j_a}$ & ( & $N_c$, & $1$, & $1$, & $\tnf$, & ${1\over N_c}$, &
$-{{(N_c+\tnf)N_f}\over{N_c \tnf}}$, &
$\tt$ &
) \nr
${\tZ_j^a}$ & ( & $\overline{N_c}$, &
$1$, & $1$, & $\overline{\tnf}$, & $-{1\over N_c}$, &
${{(N_c+\tnf)N_f}\over{N_c \tnf}}$, &
$\tt$ &
) \nr
$X^a_b$ & ( & $N_c^2-1$, & $1$, & $1$, & $1$, & $0$, & $0$, &
$\tt$ & ). \nr
\endtable
\noindent
The superpotential associated with these massless fields is (assuming no
quantum corrections) the same as the original superpotential \wel, since
all the fields involved remain massless. This is the only classical
interaction
between the fields in the IR.

In the ``magnetic" theory it is obvious from the form of the baryon
operator that all $N_f+\tnf$ quarks must get a VEV. This is, of course,
possible since $N_f \geq N_c$ implies $N_f+\tnf \leq 2N_f+\tnf-N_c$.
We can
always choose (by flavor and color transformations) the VEV of $q$ to
be diagonal, $q^i_j=\delta^i_j A_i$. The $F$-term equations force us
to choose the non--zero components of $z$ (by flavor and color
transformations)
to be of the form $z_j^{N_f+j} = z_j$ for $j=1,\cdots,\tnf$.
Then, the simplest choice
for the VEV of $Y$ leading to a non--zero baryon is
$Y^{N_f+\tnf+i}_{N_c+i}=y_i$ for $i=1,\cdots,N_f-N_c$ with all other
components of $Y$ vanishing.
The gauge group is completely broken
(the $q$ and $z$ VEVs break it to $SU(N_f-N_c)$, which the $Y$ VEV then breaks
completely), and the $D$ terms in the scalar potential lead to
constraints which determine (up to an overall constant) the VEVs $A_i$, $z_i$
and $y_i$. The solution turns out to be $A_i = A$ for $i=1,\cdots,N_c$,
$A_i = B$ for $i=N_c+1,\cdots,N_f$, $z_i=z$ for $i=1,\cdots,\tnf$
and $y_i=y$ for $i=1,\cdots,N_f-N_c$,
with three linear equations relating
$|A|^2$,$|B|^2$,$|z|^2$ and $|y|^2$.
Thus, we have found the VEVs of all fields
along this flat direction (all other VEVs vanish).

The VEV of $q$ breaks
the $q$-flavor $SU(N_f)$ and color $SU(2N_f+\tnf-N_c)$ symmetries together to
$SU(N_c)\times SU(N_f-N_c) \times SU(N_f+\tnf-N_c)$ where the first two
factors are global symmetries which are diagonal products of the flavor
and color symmetries, and the last factor is a local symmetry. The VEV
of $z$ breaks the local $SU(N_f+\tnf-N_c)$ symmetry to a global
$SU(\tnf)$ (which is a diagonal combination of the original $SU(\tnf)$
flavor group and part of the color group) and a local $SU(N_f-N_c)$.
The $Y$
VEV breaks the $SU(N_f-N_c) \times SU(N_f-N_c)$ symmetry to its
diagonal subgroup which remains a global symmetry. The $SU(N_f)$
$\tq$-flavor symmetry still remains, and thus we are left with the
same global symmetry as we found for the ``electric" theory \globsymb.

Next, we should check which fields become massive along this flat
direction. All of the terms in the superpotential \nwmag\ directly
give a mass to some of the fields when plugging in the VEVs.
To analyze the effect of these terms on $Y$ it is simplest
to decompose the matrix $Y$ as follows :
\eqn\ydecomp{Y = \pmatrix{ Y_1 & Y_2 & Y_3 & Y_4 \cr Y_5 & Y_6 & Y_7 & Y_8 \cr
Y_9 & Y_{10} & Y_{11} & Y_{12} \cr Y_{13} & Y_{14} & Y_{15} & Y_{16} \cr}}
where the $2N_f+\tnf-N_c$ rows and columns of the matrix $Y$
are decomposed into groups of size
$(N_c,N_f-N_c,\tnf,N_f-N_c)$. We chose a VEV in which $Y_8$ is diagonal
and the VEVs of all the other submatrices vanish.
It is then easy to see that the fields which get a mass from the superpotential
are
all the components of
$\tq$ and $\tz$; $M^i_{\ti}$ for $i=N_c+1,\cdots,N_f$; $N^i_{\ti}$;
$M^j_{\ti}$; $M^i_j$ for $i=N_c+1,\cdots,N_f$; $Y_2$; $Y_6+Y_{16}$;
a linear combination of $Y_9$ and $M^i_j$ for $i=1,\cdots,N_c$;
$Y_{10}$; $Y_{11}$;
$Y_{12}$; $Y_{13}$;
$Y_{14}$; and $Y_{15}$.
Some of the fields which remain massless ($(2N_f+\tnf-N_c)^2-1$ of them) are
swallowed by the Higgs mechanism and, as in the ``electric" theory,
one field remains massless and labels the flat direction.
These can be seen to be all the components of $q$ (including
$N_f(2N_f+\tnf-N_c)$ superfields), all the components of $z$
(including $\tnf(2N_f+\tnf-N_c)$ superfields)
and the matrices $Y_5$,$Y_6-Y_{16}$,$Y_7$ and $Y_8$,
which include $(N_f-N_c)(2N_f+\tnf-N_c)$ more superfields.

The fields which remain massless are thus
$Y_4$;
$M^i_{\ti}$ for $i=1,\cdots,N_c$;
$Y_3$; a linear combination of $Y_9$ and $M^i_j$
for $i=1,\cdots,N_c$; and $Y_1$.
By computing the global quantum numbers of all these fields, one can find
that they are exactly the
same as those we found above for the ``electric" fields, if we
identify them (in the order in which they appear above) with the ``electric"
fields which remained massless (in the order they appear in the table
which gives their quantum numbers).
The superpotential of the massless fields turns
out to be, after integrating out the massive fields,
exactly the same as the one we have found in the ``electric" theory, with the
appropriate identification of the fields.
The only contribution is from the $\Tr(Y^3)$ term acting on the submatrix
\eqn\yhat{{\hat Y} = \pmatrix{ Y_1 & Y_3 \cr Y_9 & Y_{11} \cr}.}

Thus, we find, as in Seiberg's duality, that along this
baryonic flat direction the ``electric" and ``magnetic" theories go
over to the same IR theory, providing additional
support for the duality
conjecture.

\subsec{The adjoint flat direction -- flow to Seiberg's duality}

Another field which can sometimes get a VEV along a flat direction is the
field $\Tr(X^2)$ (or $\Tr(Y^2)$ in the ``magnetic" theory). Since $\Tr(X)=0$
and the equation of motion of $X$ (in the absence of VEVs for the $Z$-quarks)
forces $X^2$ to be proportional to the identity matrix, this is only possible
when $N_c$ is even. In this case we may have a flat direction for which
$X^i_i=A$ ($i=1,\cdots,N_c/2$) and $X^i_i=-A$ ($i=N_c/2+1,\cdots,N_c$). Along
this flat direction the gauge group breaks to $SU(N_c/2)\times
SU(N_c/2)\times U(1)$, and the $Z$-quarks all become massive. Thus,
we flow to two SQCD theories with $N_f$ flavors and $N_c/2$ colors,
coupled by a $U(1)$ gauge symmetry. The fields corresponding to the
diagonal submatrices of $X$
become massive, while those corresponding to the
off--diagonal submatrices of $X$
remain massless and are swallowed by the Higgs mechanism.

In the ``magnetic" theory, according to \trident, $Y$, $z$, and $\tz$ must
all get VEVs along this flat direction (since the operator identified with
$\Tr(X^2)$ must be non--zero while the operator identified with $Z^j \tZ_j$
must be zero). In fact, $z$ and $\tz$ must get diagonal VEVs, because we
assume that
the adjoint $z$-mesons do not get VEVs. Examining the equations of motion
and the tracelessness condition of $Y$ we find that this is only possible
for even $2N_f-N_c$.
Then, $Y$ can get a VEV of the
form $Y^i_i=A$ ($i=1,\cdots,N_f-N_c/2$) and
$Y^i_i=-A$ ($i=N_f-N_c/2+1,\cdots,2N_f-N_c$),
where $A$ is proportional to the VEV
of $z$ and $\tz$. The gauge group breaks to $SU(N_f-N_c/2)\times SU(N_f-N_c/2)
\times U(1)$, and the $z$-quarks are swallowed by the Higgs mechanism (or
get a mass from the superpotential). We, therefore, flow to two ``magnetic"
SQCD theories with $N_f$ flavors and $N_f-N_c/2$ colors.
The fields which were contained in $Y$ either get a mass from the
superpotential or remain massless and are swallowed by the Higgs mechanism.
There
is also an additional $U(1)$ local symmetry which couples the two theories.
The superpotentials in the ``magnetic" theories flow to two copies of
the superpotential of Seiberg's ``magnetic" theory in the same way as
described in
the discussion of the adjoint mass perturbation in section 3.3.

Thus, we find here the same results as those that we obtained
in the investigation of the
adjoint mass perturbation (in the cases for which the quantum corrections
were not important). Again, we flow
to two copies of Seiberg's dual
theories \duals, this time with the same gauge group for the two copies, and
with an additional $U(1)$ gauge symmetry in both theories. As in the
analysis of the adjoint mass perturbation, we expect quantum effects to be
important when $N_f \leq N_c/2$. The above analysis is only relevant when $N_f
> N_c/2$.

\newsec{The mesonic flat direction}

Unlike the previous flat directions, we can prove that a flat direction
along which the meson operator gets a VEV exists in the quantum theory.
This may be done as for ordinary SQCD, by giving masses to all the fields,
computing the gluino condensate in the low--energy Yang--Mills theory, and
relating it by the Konishi anomaly
to the meson VEV. This gives the VEV of
the meson in the massive theory. By taking the masses
to zero in various ways, we can get any meson VEV of rank smaller than $N_c$.
We cannot prove in this way that no other fields get VEVs along this flat
direction, but this is a reasonable assumption, which we will justify by
discussing the possible corrections to the superpotential in the two theories.
Although, for simplicity, we shall limit
the discussion of this section to the case $\tnf=0$ corresponding to \dualk,
it can be generalized to arbitrary $\tnf$.

\subsec{The mesonic flat direction in the ``electric" theory}

The classical superpotential in the ``electric" theory of $\tnf=0$ is of the
form
\eqn\wclass{W_{el} = \Tr(X^3).}
In the quantum theory, there are two types of operators which may appear
in the superpotential without breaking any of the symmetries of the theory.
The first type of operators involves the quark fields. In order to respect
the flavor symmetries these can be of the form $\det(Q X^n \tQ)$ for some
integer $n$ (which must be larger than zero if $N_f \geq N_c$, and
non--negative otherwise). The second type of operators involves only the
field $X$, like $\Tr(X^n)$ (there is a finite number of independent operators
of this sort) or $\det(X)$. All these operators must of course appear with
appropriate powers so that the resulting R charge will be 2.
Of course, products of operators of both types may also appear.
Envisioning the
most general superpotential which one can write down using
these terms, it is hard
to see how $X=0$ will not turn out
to be a solution of the
equations of motion also
when we give a VEV to one quark and one anti--quark flavor (so as to be
along the mesonic flat direction). We assume, of course, that the point for
which all VEVs vanish is in the quantum moduli space of the theory for all
$N_f>N_c/2$. Otherwise the discussion of the duality should be changed
completely.
Of course, in the ``electric" theory, the rank of the meson VEV cannot be
larger than $N_c$.
Hence, it is reasonable to assume that the quantum theory
has a mesonic flat
direction, for which the meson is the only gauge--invariant operator getting
a VEV, at least for mesons of small rank.

Along this flat direction (assuming a meson VEV of rank
one), since one quark and one
anti--quark acquire a VEV and $X$ does not, the gauge symmetry is broken
to $SU(N_c-1)$. We will choose the fields which acquire non--zero
VEVs to be $Q^{N_f}_{N_c}$ and $\tQ_{N_f}^{N_c}$.
The fields
$Q^{N_f}$ and $\tQ_{N_f}$ (all components apart from one which labels the
flat direction) now get
swallowed by the Higgs mechanism.
The field $X$ includes an adjoint, a fundamental, an anti--fundamental
and a singlet representation of the new gauge group.
Ignoring the quantum corrections
all of these fields also remain massless.
Thus,
we obviously do not flow to the ``electric" theory with $N_f-1$ flavors
and $N_c-1$ colors, since we have additional superfields which have
non--trivial interactions (classically among themselves and with the field
$X$, and in the quantum theory perhaps with the quarks as well).
Denoting the $SU(N_c-1)$ adjoint field in $X$
by $\hat X$, the fundamental and anti--fundamental
fields by $Z$ and $\tilde Z$, and the singlet field (with an arbitrary
normalization) by $A$, one can easily find that the original superpotential
becomes
\eqn\wnew{\eqalign{W&=\Tr(X^3) \cr
&=\Tr({\hat X}^3) + 3 A \Tr({\hat X}^2) + 3 Z_a
{\hat X}^a_b
{\tilde Z}^b -3(N_c-2) A Z_a {\tilde Z}^a - (N_c-2)(N_c^2-N_c+1) A^3. \cr}}
The resulting theory has a global symmetry
\eqn\globsym{SU(N_f-1) \times SU(N_f-1) \times U(1)_{\hat B} \times U(1)_Z
\times U(1)_{\hat R},}
where $U(1)_{\hat B}$ and $U(1)_{\hat R}$ are the
new baryon number and $R$-charge symmetries (the old ones were
broken by the VEV).
There is an additional $U(1)_Z$ which remains a symmetry of
the theory. It is a combination of two $U(1)$'s originating from the two
(broken) flavor $SU(N_f)$ groups and of a $U(1)$ originating from the (broken)
color group.

In addition to the previously mentioned fields,
the quarks (and anti--quarks) whose color
index corresponds to the direction of the VEVs
also remain massless and decouple from the
theory.
The full field content of the resulting theory, with the quantum numbers of
$SU(N_c-1)$ and of the global symmetry group
$SU(N_f-1)\times SU(N_f-1) \times
U(1)_{\hat B} \times U(1)_Z \times U(1)_{\hat R}$, is :
\thicksize = 0pt
\thinsize = 0pt
\vskip 1.truecm
\begintable
$Q^i_a$ & $N_c-1$ & ( & $N_f-1$ ,&$1$,&${1\over {N_c-1}}$,&
${{N_c-N_f}\over{N_c-1}}$, &
${{N_f-\tt N_c}\over
{N_f-1}}$&
)  \nr
$ Q^i_{N_c}$ & $1$ & ( & $N_f-1$,&$1$,&$0$,&
$N_f$, &
${{N_f-\tt N_c}\over{N_f-1}}$&
) \nr
$\tQ^a_{\ti}$ & $\overline{N_c-1}$ & ( & $1$, & $\overline{N_f-1}$, &
$-{1 \over {N_c-1}}$, &
${{N_f-N_c}\over{N_c-1}}$, &
${{N_f-\tt N_c}\over {N_f-1}}$ &
) \nr
$\tQ^{N_c}_{\ti}$ & $1$ & ( & $1$, & $\overline{N_f-1}$, &
$0$, &
$-N_f$, &
${{N_f-\tt N_c}\over {N_f-1}}$ &
) \nr
${\hat X}^a_b$ & $((N_c-1)^2-1)$ & ( & $1$, & $1$, & $0$, & $0$, &
$\tt$ & ) \nr
${Z_a}$ & $N_c-1$ & ( & $1$, & $1$, & ${1\over {N_c-1}}$, &
$-{{N_c(N_f-1)}\over{N_c-1}}$, &
$\tt$ &
) \nr
${\tZ^a}$ & $\overline{N_c-1}$ &
( & $1$, & $1$, & $-{1\over {N_c-1}}$, &
${{N_c(N_f-1)}\over{N_c-1}}$, &
$\tt$ &
) \nr
$A$ & $1$ & ( & $1$, & $1$, & $0$, & $0$, & $\tt$ & ) \nr
$W_{\alpha}$ & $((N_c-1)^2-1)$ & ( & $1$, & $1$, & $0$, & $0$, & $1$ & ). \nr
\endtable
\noindent
The $R$ charge pertains to the lowest component of each superfield, and the
superpotential is given by \wnew.

\subsec{The mesonic flat direction in the ``magnetic" theory}

In the ``magnetic" theory of $\tnf=0$ \dualk\
the classical superpotential is of the form
\eqn\wclassm{W_{mag} = M^i_{\ti} q_i Y \tq^{\ti} + N^i_{\ti} q_i \tq^{\ti} +
\Tr(Y^3).}
For this superpotential it is obvious that any meson VEV, with zero
VEVs for the
other fields, is a flat direction. As in Seiberg's ``magnetic" theory
\duals, this superpotential
(at least for $N_f > N_c$)
is subject to quantum corrections. These are necessary in order to match
the ``electric" theory, in which a meson cannot get a VEV of rank higher than
$N_c$.
Possible
terms in the full quantum superpotential which include meson operators
are for instance $\det(M)$
or $M q Y^n \tq$, raised to appropriate powers so as to have the correct R
charge.
However, it seems that any meson VEV of small rank,
with zero VEVs for all other fields, must be a solution of the equations
of motion for the most general superpotential which one can write.
For instance, the equation of motion of the possible term in the superpotential
of the form
$\det(M)^{1/(N_f-{2\over 3}N_c)}$ vanishes whenever the rank of
$M$ is less than ${2\over 3} N_c$. Thus, it seems (though again we have
no rigorous proof of this) that also in the ``magnetic" theory there is a
flat direction, for which $M$ is non--zero of small rank (in particular of
rank 1 as we want) while all other operators vanish.

The effect of going far along this flat direction is less clear than in
the ``electric" theory. In Seiberg's ``magnetic" theory \duals,
a VEV to $M$ was
equivalent to a mass term for one of the dual quarks. Thus, going along
this flat direction we integrated out the quark that became massive.
However, here $M$ is the coefficient of $q Y \tq$, so it is not clear what
is the meaning of taking one component of $M$ to be very large. The
problem arises because the relevant term in the superpotential is
non--renormalizable. We want to take $M^{N_f}_{N_f}$ to be very large,
and, therefore, the high--energy action which leads to the
generation of this non--renormalizable term becomes important. We will
proceed by choosing a particular high--energy renormalizable action which
leads to \wclassm\ upon integrating out the massive fields. The
choice is not unique, but our choice seems to be the simplest possible
one. We shall later be able to check that the resulting low--energy
theory is indeed dual to the theory we found along this flat direction in
the ``electric" theory.

To define
the high--energy theory we introduce new massive superfields $\hW^{\ti}_a$,
$W^a_{\ti}$, $\htW_i^b$ and $\tW^i_b$, where $i$ and $\ti$ are flavor
indices of the $SU(N_f)$ flavor groups and $a$ and $b$ are color indices of
$SU(2N_f-N_c)$. We then write the superpotential
\eqn\wmass{
W_{UV} = M^i_{\ti} q^a_i \hW^{\ti}_a + Y^b_a \tq^{\ti}_b
W^a_{\ti} + M_W \hW^{\ti}_a W^a_{\ti} +
M^i_{\ti} \tq^{\ti}_b \htW^b_i + Y^b_a q^a_i \tW^i_b +
M_W \htW^b_i \tW^i_b.}
This superpotential does not break any of the symmetries of the theory
(upon giving the $W$ fields appropriate charges). Upon integrating out
the massive $W$ fields it leads to the term $M^i_{\ti} q^a_i Y^b_a
\tq^{\ti}_b$ in the low--energy superpotential. Now, in this high--energy
theory, a non--zero $M^{N_f}_{N_f}$ is still a flat direction of the
theory, and we can analyze what happens when we take it to be large
(of the same order of magnitude as $M_W$).
We find that the field proportional to $M_W W_{N_f} + M^{N_f}_{N_f} q_{N_f}$
gets a mass instead of $W_{N_f}$, as does the field proportional to
$M_W {\tilde W}^{N_f} + M^{N_f}_{N_f} \tq^{N_f}$, while the orthogonal
combinations remain massless. We will denote these massless fields by
$z^a$ and $\tz_a$, and up to a normalization constant they are given by
\eqn\massless{\eqalign{
z^a &= -M^{N_f}_{N_f} W_{N_f}^a + M_W q_{N_f}^a \cr
\tz_a &= -M^{N_f}_{N_f} \tW^{N_f}_a + M_W \tq^{N_f}_a. \cr }}
We can now integrate out all the massive fields (using
the full superpotential which includes also the other terms in \wclassm),
and find the superpotential for the massless fields.
This superpotential includes terms like \wclassm\ but
involving only the first $N_f-1$ flavors. In addition to this there are
various terms involving the fields of the $N_f$'th flavor. The renormalizable
terms are proportional to $z^a Y^b_a \tz_b$, $N^{N_f}_{\ti} \tq^{\ti}_a
z^a$, $N_{N_f}^i q_i^a \tz_a$ and $N^{N_f}_{N_f} z^a {\tz}_a$, and
there are additional non--renormalizable terms, involving also the fields
$M_{N_f}^i$ and $M^{N_f}_{\ti}$. Similar results may be found from other
high--energy superpotentials which give the same low--energy superpotential.
The differences are generally only in irrelevant non--renormalizable terms.
$M^{N_f}_{N_f}$ is also massless and labels the flat direction, as does
one of the components of $Q^{N_f}$ and $\tQ_{N_f}$ in the ``electric"
theory.

As in the ``electric" theory, we find that the remaining global symmetry
is of the form $SU(N_f-1)\times SU(N_f-1) \times U(1)_{\hat B} \times
U(1)_Z \times U(1)_{\hat R}$. The $U(1)_Z$ symmetry here is the sum
of two $U(1)$ factors from the (broken) $SU(N_f)$ flavor groups.
The quantum numbers of the remaining massless
fields under the (unbroken)
$SU(2N_f-N_c)$ color group and under the flavor group
can easily be seen to be (where we choose the normalization of the
new $U(1)_Z$ to agree with its normalization in the ``electric" theory) :
\thicksize = 0pt
\thinsize = 0pt
\vskip 1.truecm
\begintable
$q_i^a$ & $2N_f-N_c$ & ( & $\overline{N_f-1}$, & $1$, &
${1\over{2N_f-N_c}}$, &
$-1$, &
$1-{1\over 3}{{4N_f-2N_c-1}\over{N_f-1}}$ &
) \nr
${\tq}_a^{\ti}$ & $\overline{2N_f-N_c}$ & ( & $1$, & $N_f-1$, &
$-{1\over{2N_f-N_c}}$, &
$1$, &
$1-{1\over 3}{{4N_f-2N_c-1}\over{N_f-1}}$ &
) \nr
$Y$ & $((2N_f-N_c)^2-1)$ & ( & $1$, & $1$, & $0$, & $0$, & $\tt$ & ) \nr
$M^i_{\ti}$ & $1$ & ( & $N_f-1$, & $\overline{N_f-1}$, & $0$, &
$0$, &
$2{{N_f-\tt N_c}\over{N_f-1}}$ &
) \nr
$N^i_{\ti}$ & $1$ & ( & $N_f-1$, & $\overline{N_f-1}$, & $0$, &
$0$, &
$\tt+2{{N_f-\tt N_c}\over{N_f-1}}$ &
) \nr
$M^{N_f}_{\ti}$ & $1$ & ( & $1$, & $\overline{N_f-1}$, & $0$, &
$-N_f$, &
${{N_f-\tt N_c}\over{N_f-1}}$ &
) \nr
$N^{N_f}_{\ti}$ & $1$ & ( & $1$, & $\overline{N_f-1}$, & $0$, &
$-N_f$, &
$\tt+{{N_f-\tt N_c}\over{N_f-1}}$ &
) \nr
$M_{N_f}^i$ & $1$ & ( & $N_f-1$, & $1$, & $0$, &
$N_f$, &
${{N_f-\tt N_c}\over{N_f-1}}$ &
) \nr
$N_{N_f}^i$ & $1$ & ( & $N_f-1$, & $1$, & $0$, &
$N_f$, &
$\tt+{{N_f-\tt N_c}\over{N_f-1}}$ &
) \nr
$N_{N_f}^{N_f}$ & $1$ & ( & $1$, & $1$, & $0$, &
$0$, & $\tt$ & ) \nr
$z^a$ & $2N_f-N_c$ & ( & $1$, & $1$, & ${1\over {2N_f-N_c}}$, &
$N_f-1$, &
$\tt$ &
) \nr
$\tz_a$ & $\overline{2N_f-N_c}$ &
( & $1$, & $1$, & $-{1\over {2N_f-N_c}}$, &
$-(N_f-1)$, &
$\tt$ &
) \nr
$W_{\alpha}$ & $((2N_f-N_c)^2-1)$ & ( & $1$, & $1$, & $0$, & $0$, & $1$ & ).
\nr
\endtable
\noindent

\subsec{Comparison of the resulting theories}

Now that we have found the resulting theories along the mesonic flat direction,
we can compare them to see that they are indeed dual.
First, we notice that all singlets appearing in the ``electric" theory have
partners in the ``magnetic" theory with the same quantum numbers,
and, therefore,
we can identify them. Next, we note that without the singlets, the
``electric" theory is exactly the ``electric" theory of section 2 with
$N_f-1$ regular flavors, $\tnf=1$ and $N_c-1$ colors. The ``magnetic"
theory is exactly the ``magnetic" theory of section 2 with $N_f-1$
regular flavors, $\tnf=1$ and $2N_f-N_c=2(N_f-1)+1-(N_c-1)$ colors.
Thus, these theories are dual to each other. By giving a mass
to all the ``electric" singlets we can flow directly to the dual
theories described in section 2. In fact,
by taking a meson VEV of rank $k$ (with $k$ equal non--zero eigenvalues)
we can flow in this way to dual theories
with $N_f-k$ regular
quark flavors, $\tnf=k$ and $N_c-k$ colors. Starting from a theory with general
$N_f$,$\tnf$ and $N_c$ we would find that $N_f$ and $N_c$ decrease by $k$
while $\tnf$ increases by $k$, with no change in the ``magnetic" gauge
group, in a way consistent with the duality.
All the duality theories we analyze may
be obtained in this way from the $\tnf=0$ theory of \dualk,
and this is in fact how
we first discovered them.

\newsec{Summary and conclusions}

In this paper we described a duality transformation of $N=1$
supersymmetric non--abelian gauge theories. This duality can be viewed as
resulting from Kutasov's duality \dualk\ by flowing along one of its flat
directions. We have performed several
non--trivial checks for the consistency of this duality.
These include mass perturbations
and the investigation of the theories along several flat
directions.

To further analyze these
theories one should understand the quantum corrections to their superpotential.
This is necessary for the identification of all flat directions of the two
theories. We recall that in some of the cases investigated in this work,
and certainly in other cases as well,
the quantum corrections are important. The quantum corrections may be
similar in form to the quantum superpotential recently computed in \efgr.
It may also be interesting to generalize the duality transformation
analyzed here to
other gauge groups, such as $SO(N)$.

By flows along flat directions or mass perturbations we can connect our
dual theories with those of Seiberg \duals\ and Kutasov \dualk. Thus, it
seems clear that the (as yet not well understood) mechanism behind the
duality symmetry is the same for all of these cases. These include all
cases which are known so far of
$N=1$ $SU(N)$ gauge theories which exhibit duality.
Understanding this mechanism in
one of the cases will, therefore, enable us to understand all of them.
In particular, the case of $N_f=0$ and $\tnf=2N_c$ seems to be related
to the duality of $N=2$ theories with $N_f=2N_c$ \refs{\swi-\klyt}.
{}From this theory we can flow to all other theories with $N_f=0$, but not to
theories with $N_f>0$. The relation between the duality of section 2
and the $N=2$
duality is under current investigation. Such a relation would also relate
the $N=2$ duality to the dualities of Seiberg \duals\ and Kutasov \dualk.

Another interesting generalization of our results is to duality symmetries
between other $N=1$ gauge theories, in particular theories with other
matter representations. We have noticed that the 't-Hooft anomaly
matching conditions associated with the original $SU(N_c)$ ``electric" theory
of SQCD are
also satisfied by a class of ``magnetic" theories, generalizing the
solution found in \duals. The ``magnetic"
theories are $SU(\tilde N_c)$ gauge theories
with $\tilde N_c=kN_f-N_c$, with $k$
underlying mesons having $U(1)_R$ charges differing by multiples of 2
starting with $2-2{N_c\over N_f}$ (for the scalar meson),
and with the same dual quarks and gluinos.
Moreover, we have found, for these cases, a complete identification
between the gauge--invariant
operators in the ``electric" and ``magnetic" theories.
It would be interesting to find out whether one can ``elevate" these
cases to a full duality symmetry.
We hope to return to the various possible generalizations in future work.
Clearly, the phenomenon of $N=1$ duality is quite general. Hopefully,
the investigation of these generalizations will shed light on the underlying
mechanism behind the duality symmetry.

\listrefs

\end